\documentclass[english,a4paper,14pt]{article}
\usepackage{babel,amssymb,hyperref}
\usepackage[dvips]{graphicx}
\usepackage{epsfig}
\usepackage{afterpage}
\makeatletter
\renewcommand{\@cite}[2]{({#1\if@tempswa , #2\fi})}
\renewcommand{\@biblabel}[1]{\hfill}
 \makeatother

\advance\textheight \topskip \textwidth 35.5pc \oddsidemargin 20pt
\begin{document}

\author{M. S. Pshirkov(1)\footnote{E-mail:pshirkov@prao.ru}, M. V. Sazhin(2)}

\title{\bf{Weak microlensing effect and stability of pulsar time scale}}

\date{\emph{\footnotesize{(1) Pushchino Radio Astronomy Observatory, Astro Space Center, Lebedev Physical Institute, Pushchino,
Russia\\(2) Sternberg Astronomical Institute, Moscow State
University, Moscow, Russia}}} \maketitle
\begin{abstract}
An influence of the weak microlensing effect on the pulsar timing
is investigated for pulsar B1937+21. Average residuals of Time of
Arrival (TOA) due to the effect would be as  large as 10 ns in 20
years observation span. These residuals can be much greater (up to
1 ms in 20 years span) if pulsar is located in globular cluster
(or behind it).
\end{abstract}

\section{Introduction}
First, the problem of the influence of the weak microlensing
effect on the pulsar timing observations was discussed in
\cite{saz1986}. It was considered as interstellar Shapiro effect.
The massive body that flies not far from the line pulsar-observer
produces changes in the observing frequency of the pulsar similar
to glitches. Estimations were made for Crab and Vela pulsars,
glitches in these pulsars can be partially explained by the
influence of the effect. Substantial contribution to the problem
was made by  \cite{ld1995}; they mainly investigated the case of
microlensing (i.e. the gravitational deflector flied very close to
the  line  observer-pulsar). It was shown that the microlensing
effect would cause short-term growth of the residuals and
follow-up relaxation. Whole interaction would take less then
several years and the maximum amplitude of the residuals would be
~20-30 ms.  Such remarkable events are very rare, but all the
pulsars are affected by the weak microlensing effect to a greater
or lesser extent. This effect was considered in \cite{ohn1996},
where timing of  millisecond pulsars was proposed as detection
method for MACHOs. Growth of number of observed pulsars and time
span of observation would make such detection easier. Numerical
estimates were made in \cite{hosokawa1999}. They stated that even
when the measurement accuracy reaches to 10 ns, probability of the
remarkable influence would be in the order of $10^{-1}$ for the
pulsar of a few kpc distance from us observed over ten years.
 On the other hand there's well developed formalism for the effect that came from the optics.
 The weak microlensing effect causes distant sources like quasars from ICRF  to "tremble" on the level of tens of mas. It was  shown in \cite{sazh1996,sazh1998} that these angular fluctuations range from a few up to hundreds of microarcseconds and this leads to a small rotation of the celestial reference frame. In  \cite{sazh2001} influence of the effect on parallax measurements was considered-apparent parallax can be even negative due to the influence of the effect. Also, the weak microlensing effect can affect VLBI observations \cite{sazh2005} and it should be taken into account with new generation of space-based VLBI. In \cite{kalinina2006} some statistical studies with toy-models were made, that was applied later to real model of the
 Galaxy. In fact, both weak microlensing effect and fly-by effect
 on timing are very similar and can be considered as manifestation
 of 4D (four-dimensional) astrometry \cite{ilyasov1990}

In this work we tried to apply eikonal formalism that was
developed earlier for investigation of weak microlensing effect
for  use in pulsar timing studies.

The paper is organized as follows. In Section 2 we give a short
review of influence of a passing body on pulsar timing in eikonal
approximation. In Section 3 we apply a model of distribution of
stars in our Galaxy to numerical estimations of their influence on
pulsar timing and conclude our consideration in Section 4.

\section{Eikonal formalism applied to pulsar timing}
Change of phase during the propagation of electromagnetic wave can
be obtained as a solution of Hamilton-Jacobi equation for a
massless particle:
 \begin{equation}
\label{eik1} g^{\mu\nu}
\frac{\partial{S}}{\partial{x^{\mu}}}\frac{\partial{S}}{\partial{x^{\nu}}}=0
\end{equation}
Though  $S$ formally   is a function of action, we hereafter
identify it as eikonal or wave phase along the trajectory of the
ray of light.

In weak field approximation the metric tensor of gravitational
field can be written down in a following form
 \begin{equation}
\label{eik2}
 g_{\mu\nu}= \eta_{\mu\nu}+ h_{\mu\nu}
\end{equation}
Here $\eta_{\mu\nu}$ -- is flat Minkowskian metric,  $h_{\mu\nu}$
-- small additions to the flat metric that describes gravitational
field of spherically symmetric body (star)

Equation (\ref{eik1}) can be solved in the following form: we take
an exact solution \cite{weinberg} and then take its asymptotic
when the impact parameter of the propagating ray is much larger
then the Schwarzschild radius $r_g=\frac{2GM}{c^2}$  ($M$-mass of
the deflector)
\begin{equation}
\label{eik3} \psi=\psi_l+\frac{r_g\omega}{c}arch(\frac{r}{\rho})
\end{equation}
Here    $\psi$ is full change of the phase along the trajectory,
$\psi_l$- change of the phase along the trajectory that
corresponds to the propagation in the flat space and time, $r_g$ -
Schwarzschild radius of the deflector, $\omega$-frequency of the
electromagnetic wave, r- some point  on the trajectory, $\rho$ -
impact parameter (i.e. minimal distance between deflector $D$ and
curve of  photon propagation).

\begin{figure}[ht]
\renewcommand{\textfraction}{\0.15}
\renewcommand{\topfraction}{\0.85}
\renewcommand{\bottomfraction}{\0.15}
\renewcommand{\floatpagefraction}{\0.60}
\centering
\includegraphics[height=0.5\textheight]{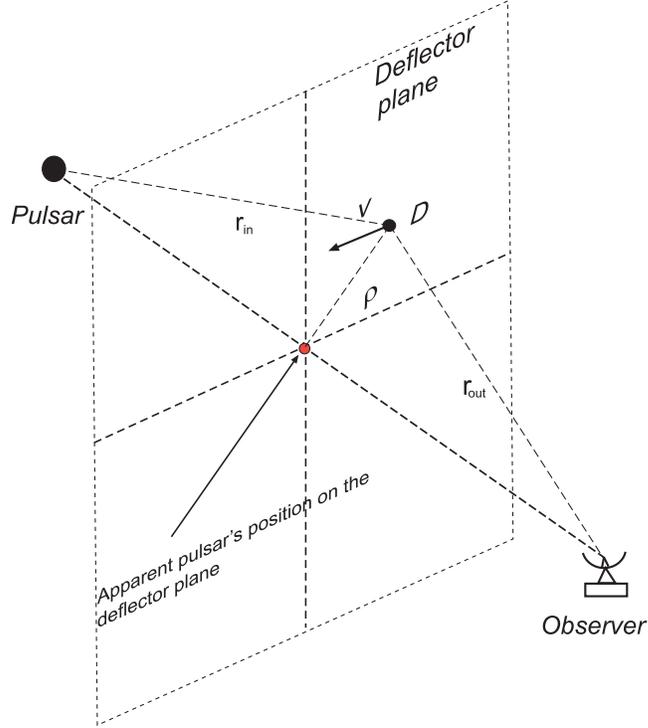}
\caption{Schematic  picture  of origin of  the weak microlensing
effect. $D$-deflecting object.
 \hfill}
\end{figure}

Only the second term in (\ref{eik3}) is a matter of interest to
us, though it's only a small addition to the usual change of phase
during the propagation.The complete phase shift can be obtained as
a sum of two solutions. The first is a phase shift during
propagation from the source of the electromagnetic waves (which is
located in $r_{in}$ ) to the closest approach to the deflector (we
set the point of origin to the center of the deflector):
 \begin{equation}
\label{eik4}
\delta\psi_{-}=\psi(r=r_{in})-\psi(r=\rho)=\frac{r_g\omega}{c}arch(\frac{r_{in}}{\rho})
\end{equation}
The second - is a phase shift during propagation from the closest
approach to the deflector to the observers (at $r_{out}$ ):
 \begin{equation}
\label{eik5}
\delta\psi_{-}=\psi(r=r_{out})-\psi(r=\rho)=\frac{r_g\omega}{c}arch(\frac{r_{out}}{\rho})
\end{equation}
And the total  phase shift is :

\begin{equation}
\label{eik6}
\Delta\psi=\frac{r_g\omega}{c}ln(\frac{4r_{in}r_{out}}{\rho^2})
\end{equation}

We treated the deflector as a  motionless body in this solution.
In fact, all stars, including MACHOs of our Galaxy are moving.
Approximate solution of space-time metric in the case of moving
deflector and trajectory of photon in such a variable
gravitational field was calculated in \cite{ks1999}. The metric
which originates from a moving body and small perturbations of
photon trajectory in gravitational field of this body, differ in
$(\frac{v}{c})^3$ terms from our solutions and we will omit this
difference. To describe the motion we take $\rho$ (impact
parameter) as function of $t$ only:

\begin{equation}
\label{eik7}
\begin{array}{l}
\Delta\psi_2-\Delta\psi_1=\frac{r_g\omega}{c}\ln(\frac{r_{in1}r_{out1}}{r_{in2}r_{out2}}\frac{\rho_{1}^2}{\rho_{2}^2}){}=\frac{r_g\omega}{c}\left[\ln(\frac{r_{out2}}{r_{out1}})+\ln(\frac{r_{in2}}{r_{in1}})+\ln(\frac{\rho^2_{1}}{\rho^2_{2}})
\right]
\end{array}
\end{equation}

Indices denotes values at different epochs $t_1$  and $t_2$. The
first two terms are negligibly small, so we can rewrite expression
(\ref{eik7}):
\begin{equation}
\label{eik8}
\Delta\psi_2-\Delta\psi_1=\frac{r_g\omega}{c}\ln(\frac{\rho^2_{1}}{\rho^2_{2}})
\end{equation}

Also we can write out  time dependence of $\rho(t)$:\\
$\rho(t)=\sqrt{\rho_0^2+v^2(t-t_0)^2}$,
 here $\rho_0$-minimal impact parameter, $v$- velocity of relative motion of pulsar
and deflector, $t_0$ - epoch of the closest approach. We can
rewrite the equation for the phase shift and obtain equation for
time delays or residuals of Time of Arrival (TOA). It's worth
noting, that these delays don't depend on frequency of
electromagnetic wave:

$$
\delta
T=\frac{\Delta\psi_2-\Delta\psi_1}{\omega}=\frac{r_g}{c}\ln(\frac{\rho_1^2}{\rho_2^2})=-\frac{r_g}{c}\ln(\frac{\rho_0^2+v^2(t_2-t_0)^2}{\rho_0^2+v^2(t_1-t_0)^2})
$$

\begin{figure}[htbp]
\renewcommand{\textfraction}{\0.15}
\renewcommand{\topfraction}{\0.85}
\renewcommand{\bottomfraction}{\0.15}
\renewcommand{\floatpagefraction}{\0.60}
 \centering
\includegraphics[height=0.3\textheight]{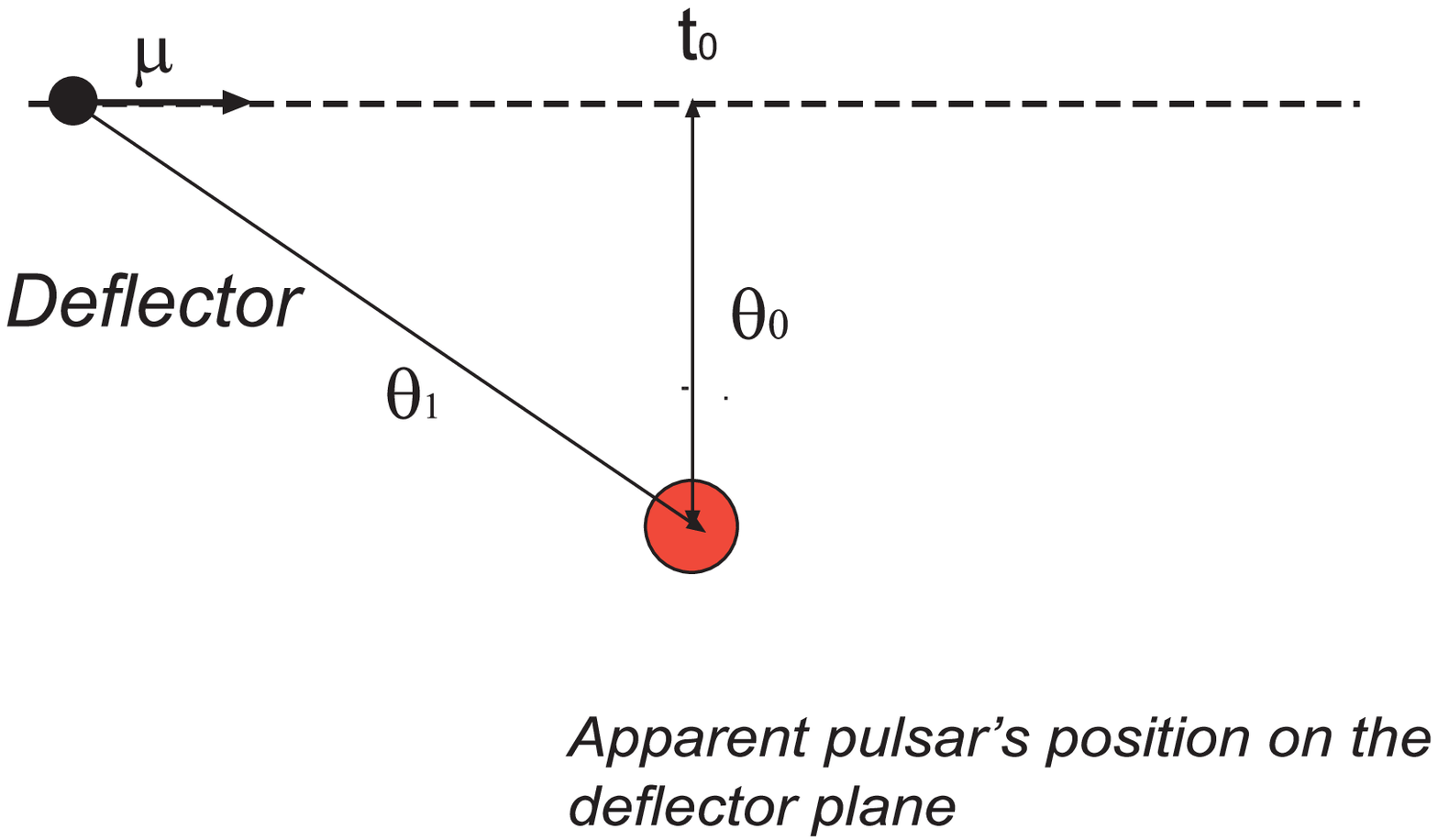}
\caption{The plane of the deflector: red circle represents
apparent position of the pulsar. Deflector with  proper motion
$\mu$ and impact parameter $\theta_0$ is passing by near this
position \hfill}

\end{figure}
We can set the first epoch $t_1$   equal to zero and discard the
second index, $t_2\equiv t$ :
\begin{equation}
\label{eik9}
 \delta T=-\frac{r_g}{c}\ln(\frac{\rho_0^2+v^2(t-t_0)^2}{\rho_0^2+v^2t_0^2})
\end{equation}
Here, $t$ is time span of observations (we set the epoch of
initial observations equal to 0), $t_0$ - is the epoch of the
closest approach of the deflector to the line of propagation.

It's convenient to consider this problem on the "plane of
deflector". Thus we convert all linear measures into angular ones:
$\rho_0=\theta_0 d$ ,$v=\mu d$, $\theta_0$  -angular distance of
the closest approach of deflector to pulsar, $\mu$ - angular
velocity of the relative motion (mainly due to the proper motion
of pulsar), $d$- distance between the deflector and the observer.
Hereafter phrases like "deflector s close to pulsar" mean we
observe close angular coincidence of the bodies, not in 3D space.
\begin{equation}
\label{eik10} \delta
T=-\frac{r_g}{c}\ln(\frac{\theta_0^2+\mu^2(t-t_0)^2}{\theta_0^2+\mu^2t_0^2})
\end{equation}

Value $\theta_0$ depends on location of pulsar in Galaxy and its
proper motion. The higher is density of deflectors in the
neighborhood of pulsar on the celestial sphere, the smaller that
value would be. We take into consideration only deflectors between
the pulsar and the observer, because they make the largest
contribution on the effect.

\section{Estimates for B1937-21}
We chose two pulsars J1643-1224 and  B1937+21 for further
estimates, because they're quite distant and located in populated
regions of our Galaxy (B1937+21:  $G_l=57.51   G_b=-0.29 r_p=3.6
kpc$; J1643-1224:  $G_l= 5.67 G_b=21.22 r_p =4.86 kpc$)
\cite{atnf1,atnf2} , so probability that effect would have place
is much higher than for other millisecond pulsars. It's essential
to define values $\theta_0$  and $t_e$ - average duration of
influence. They can be approximately found in such way
\cite{kalinina2006}: stars are nearly uniformly distributed in the
neighborhood of the pulsar on the celestial sphere; the angular
distance to the nearest star, which would affect the pulsar timing
depends on the location of pulsar.

We calculated the density of the stars in the neighborhood, using
accepted model of the disk of our Galaxy \cite{bah1986}.
\begin{equation}
\label{est1}
 N(\theta,\phi)=\int\limits_0^{r_p}n(\xi,\theta,\phi)\xi^2 d\xi
\end{equation}

$N(\theta,\phi)$ --  sought density in the direction of the
pulsar, which is assigned by the angles $\theta$,$\phi$. $\theta$-
angle between the line of sight and the Galactic plane, $\phi$-
angle between the projection of the line observer-pulsar to the
galactic plane and the line Solar system-Galactic center;$\xi$-
distance from the observer.
$$
n(r,z)=n_0\exp(-\frac{r-R_0}{3500})\exp(-\frac{z}{325}) pc^{-3}
$$
 $n_0=0.1$ -density of the stars in Sun's neighborhood, $r$ -distance from the axis of the Galaxy, z-distance from the
Galactic plane , $R_0 =8000 pc$ -distance between the Solar system
and the Galactic center, $3500 pc$ and $325 pc$ - radial and
vertical scales of the model, accordingly.

$$
r(\xi,\theta,\phi)=\sqrt{R_0^2+\xi^2\cos^2(\theta)-2R_0\xi\cos(\phi)\cos(\theta)}
$$
$$
z(\xi,\theta,\phi)=\xi\sin(\theta)
$$

\begin{equation}
\begin{array}{l}
N(\theta,\phi)=\int\limits_0^{r_p}n_0\xi^2\exp(\frac{16}{7})\\\exp(-\frac{\sqrt{R_0^2+\xi^2\cos^2(\theta)-2R_0\xi\cos(\phi)\cos(\theta)}}{3500})\exp(\frac{\xi\sin(\theta)}{325})d\xi)
\end{array}
\end{equation}

\begin{figure}[h]
\renewcommand{\textfraction}{\0.15}
\renewcommand{\topfraction}{\0.85}
\renewcommand{\bottomfraction}{\0.15}
\renewcommand{\floatpagefraction}{\0.60}
\centering
\includegraphics[height=0.5\textheight]{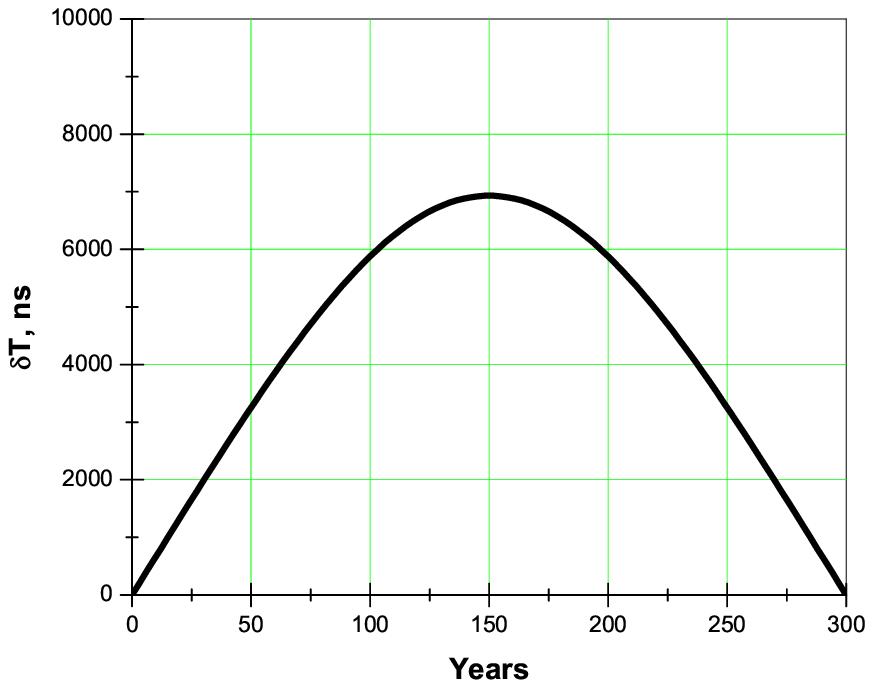}
\caption{TOA residuals, caused by the effect; no fitting conducted
yet.  \hfill}
\includegraphics[height=0.5\textheight]{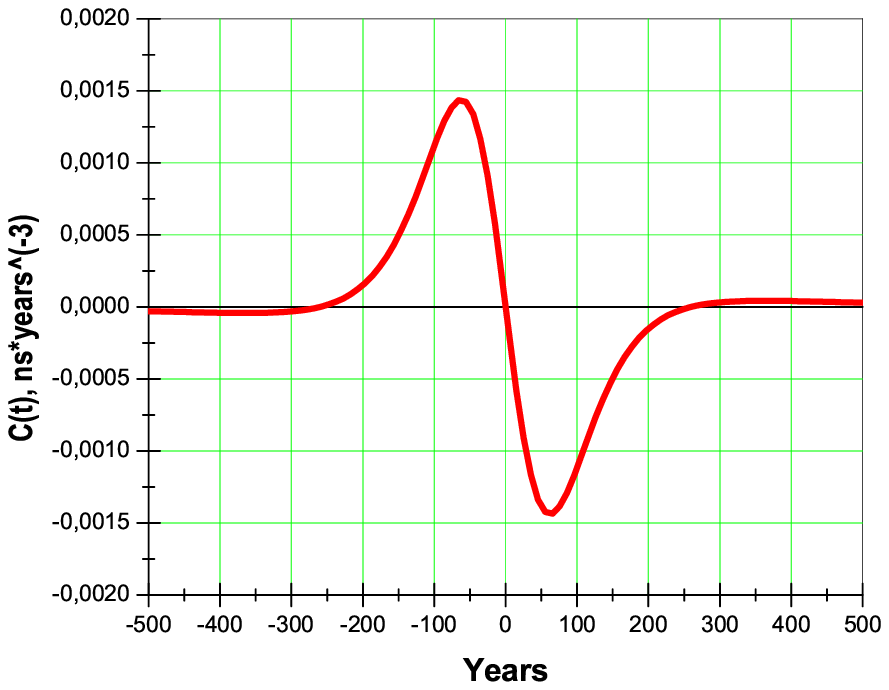}
\caption{Graph for the coefficient $C(t)$ shows that the influence
of higher-power order items should be taken into consideration.
 \hfill}
 \label{fig:c}

\end{figure}

Average angular distance $\theta_1$ between the pulsar and the
closest deflector (star) can be found with taking into account
$N(\theta,\phi)$ . Values  $\theta_0$  and  $t_e$  were calculated
using Monte-Carlo simulation: a circle of $\theta_1$ were
circumscribed around the pulsar on the celestial sphere,  then a
large amount (1000) of test deflectors with proper motion  $\mu$
were started from this circle under random angles  $\alpha$.  As a
result we found distributions for values $\theta_0$  and $t_e$,
and their averages, that were used in following estimates. Only
known distribution of stars in our Galaxy was used in our
estimates and if we take into account possible influence of Dark
Matter, then sought values can be lower in 2-3 times, because mass
of DM doesn't exceed mass of ordinary matter more than 4-5 times.
Also, we set mass of deflectors equal to $ M_{\odot}$. Values that
are essential for further estimations (J1643-1224, B1937+21) are
given in the table below.

\begin{table}[h]
\centering
\begin{tabular}{|c|c|c|c|c|}
\hline
   PSR & $\theta_1$ & $\mu$ & $\theta_0$ & $t_e$  \\
\hline
J1643-1224& 7.3''  & $\sim 10 mas/yr$&  4.7" & 470 yr \\
\hline
B1937+21 & 2.5"  & $\sim 10 mas/yr$&  1.5" & 150 yr      \\
\hline
\end{tabular}
\end{table}

\begin{figure}[ht]
\centering
\includegraphics[]{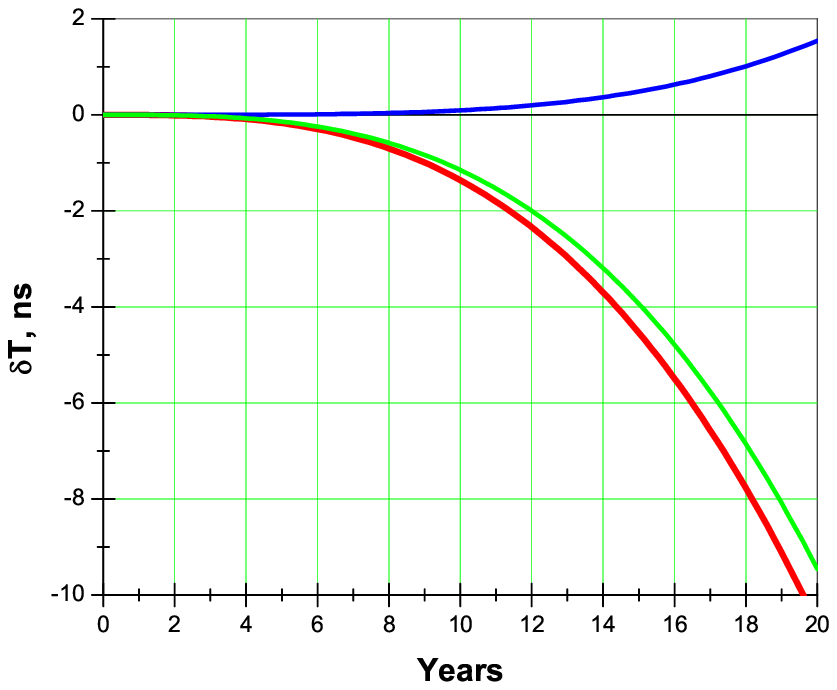}
\caption{TOA residuals due to weak microlensing effect. The blue
curve corresponds to $t_0$ =0,  green one to $t_0$=50 years and
red one to  $t_0$=100 years. \hfill} \label{fig:timing}
\includegraphics[]{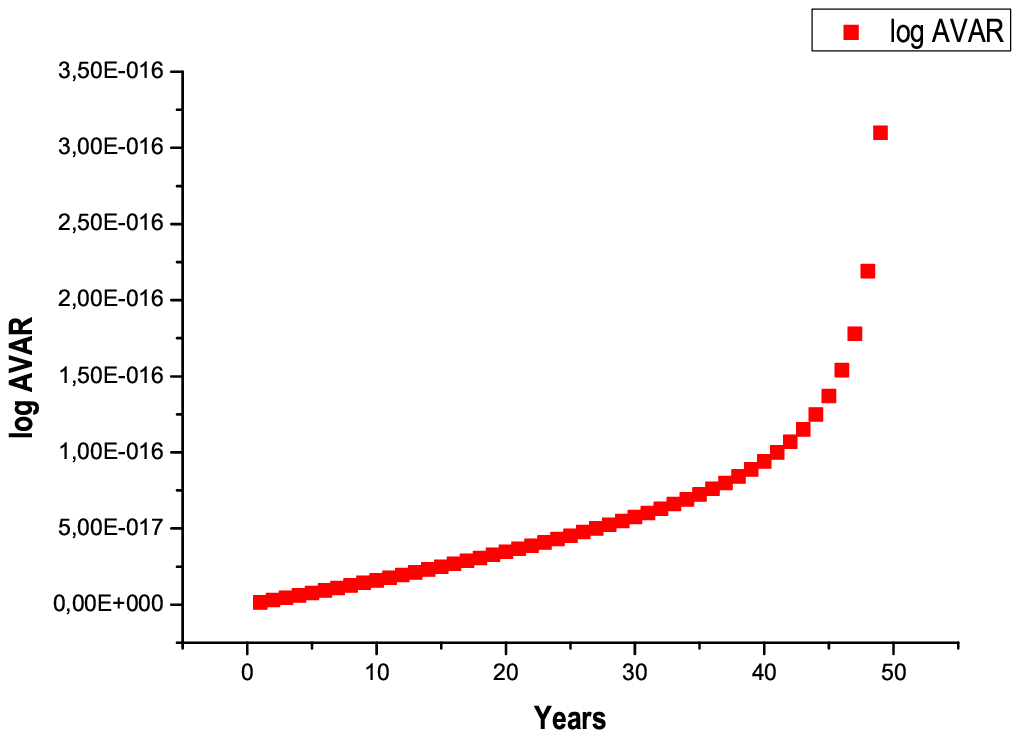}
\caption{Allan Variance (AVAR) which appears from a weak
microlensing effect.
 \hfill}

\end{figure}
We can see the influence of the effect on the residuals, but only
trends of cubic order and higher will survive during usual fitting
procedure \cite{bh1986}. Linear and quadratic terms will redefine
apparent period of pulsar $P$ and its first derivative $\dot{P}$
and can't be found.

Residuals of TOA due to the weak microlensing effect can be
written as follows:
\begin{equation}
\label{est2} \delta T_{postfit}=Ct^3+Dt^4+Et^5
\end{equation}

$C, D, E$  are coefficients in Taylor's series of function  (5)
where  $t=0$.

Plotted coefficient $C$ depending on   $t_0$ is represented in
fig.~\ref{fig:c} (plotted for B1937+21).

One can see from the plot that the fastest increase of residuals
takes place when the epoch of the initial observation are 50-150
years away from the epoch $t_0$, because the third derivative have
maximum in that interval maximal. If the initial observation
coincides with the closest approach, then only fourth and higher
orders term will affect timing and the residuals will increase
much slowly. Magnitude of the residuals after subtraction of
linear and quadratic terms can be expressed as follows: $\delta
T_{postfit}=-\frac{r_g}{c}\ln(\frac{\theta_0^2+\mu^2(t-t_0)^2}{\theta_0^2+\mu
t_0^2 })-A(0)t-B(0)t^2$, where $A(0)$, $B(0)$ -linear and
quadratic coefficients at $t=0$.

$$A(0)=\frac{2r_g}{c}\frac{\mu^2t_0}{\theta^2+\mu^2t_0^2}$$
$$B(0)=-\frac{r_g}{c}\frac{\mu^2}{\theta^2+\mu^2t_0^2}+\frac{2r_g}{c}\frac{\mu^4t_0^2}{(\theta^2+\mu^2t_0^2)^2}$$

The plot in fig.~\ref{fig:timing} shows magnitude of the residuals
at different $t_0$ (0, 50, 100 years; blue, green and red graphs
accordingly). Module of that magnitude depends only on module
$t_0$.
\begin{figure}[t]
\centering
\includegraphics[height=0.5\textheight]{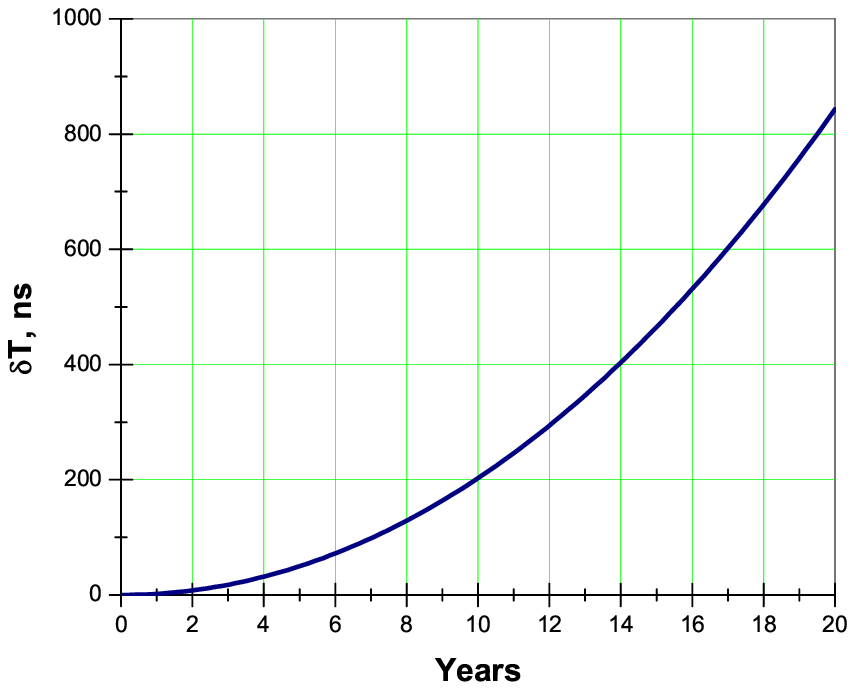}
\caption{ Strong influence of the effect. TOA residuals can reach
1 ms in 20 years observations span. \hfill}

\end{figure}

%
%

Residuals of 10 ns magnitude due to the effect of weak
microlensing  will appear with probability of  $\sim 50\%$ if time
span of observations  exceeds 20 years.

We can also calculate Allan variance (AVAR) for pulsar time scale
with time residuals caused  by the effect.

TOA residuals due to the effect can be significant, if  $\theta_1$
(angular distance between the pulsar and the nearest affecting
body) is much smaller than average. The plot   Fig.7 represents
situation when $\theta_0=0.1 mas$. This situation has $\sim 0.5\%$
chance of probability in case of B1937+21; Probability reduces
like $\theta_0$ inverse squared. The magnitude of the residuals
can be as a great as 800-1000 ns in the same 20 years span.

However, if we used in fitting procedure terms of cubic and higher
orders, then the magnitude of the effect can be effectively set to
0.

The magnitude can be much greater for pulsars in GC (or pulsars
behind GC) ($n=10^{3-4} pc^{-3}$, $L$ (length of path of ray in
GC)$=10 pc$, $d$(distance to GC) = $1-10 kpc$)). $t_e$  and
$\theta_0$ can be much smaller because the density of stars  in GC
is large, The magnitude of the effect will be much greater (~the
same 1 ms in 20 years span) . Time of one significant interaction
will be quite small (20-30 years). Complete investigation of the
question can be found in \cite{sazh1993,lk2006}.
%
\section{Conclusions}
So, we can make several conclusions: average TOA residuals due to
a weak microlensing effect is about 10 ns (B1937+21) in 20 years
span. TOA residuals can be effectively set to zero by using higher
order terms in fitting procedure (not for pulsars in globular
clusters. Residuals   can be   much greater if pulsar is located
in a globular cluster, so the pulsars in globular clusters can't
be recommended for using in PT scale.


\end{document}